\documentclass{article}
\usepackage{spconf, amsmath, amssymb, epsfig, subcaption, bm, multirow}

\title{Model-based demosaicking for acquisitions\\by a RGBW color filter array}
\name{Matthieu Muller$^{\dagger}$ \qquad
      Daniele Picone$^{\dagger}$ \qquad
      Mauro Dalla Mura$^{\dagger \ddagger}$ \qquad
      Magnus O. Ulfarsson$^{\star}$}
  
\address{$^{\dagger}$ Grenoble Alpes University, CNRS, Grenoble INP, GIPSA-lab, 38000 Grenoble, France \\
         $^{\ddagger}$ Institut Universitaire de France (IUF), France \\
         $^{\star}$ University of Iceland, Faculty of Electrical and Computer Engineering, 101 Reykjavik, Iceland}
\begin{document}
%
\maketitle
\begin{abstract}
Microsatellites and drones are often equipped with digital cameras whose sensing system is based on color filter arrays (CFAs), which define a pattern of color filter overlaid over the focal plane. Recent commercial cameras have started implementing RGBW patterns, which include some filters with a wideband spectral response together with the more classical RGB ones. This allows for additional light energy to be captured by the relevant pixels and increases the overall SNR of the acquisition. Demosaicking defines reconstructing a multi-spectral image from the raw image and recovering the full color components for all pixels. However, this operation is often tailored for the most widespread patterns, such as the Bayer pattern. Consequently, less common patterns that are still employed in commercial cameras are often neglected. In this work, we present a generalized framework to represent the image formation model of such cameras. This model is then exploited by our proposed demosaicking algorithm to reconstruct the datacube of interest with a Bayesian approach, using a total variation regularizer as prior. Some preliminary experimental results are also presented, which apply to the reconstruction of acquisitions of various RGBW cameras.
\end{abstract}
\section{Introduction}
\label{sec:intro}

In certain remote sensing setups, constellations of microsatellites or drones are utilized for acquiring data. These platforms have limited payload capacity, which necessitate the use of novel solutions for the onboard instruments and often incorporate miniaturized commercial cameras.

In this configuration, the acquired raw image is transmitted as a downstream data to a ground station, which is responsible for processing it to reconstruct the desired datacube. Cameras based on color filter arrays (CFAs) represent a popular choice for this task. The CFA defines an array of spectral filters overlaid over the sensors, typically arranged in periodic patterns. The incident light is consequently filtered by their unique spectral response of each filter, before being captured by the sensor.

\begin{figure}[!t]

\begin{subfigure}{.475\linewidth}
  \centerline{\includegraphics[width=.75\linewidth]{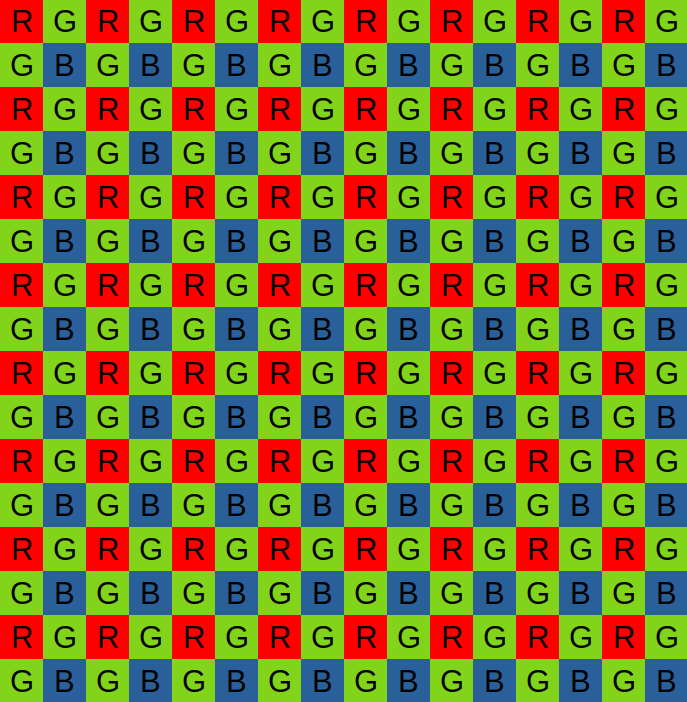}}
  \caption{Bayer CFA}
  \label{fig:bayer}
\end{subfigure}\hfill 
\begin{subfigure}{.475\linewidth}
  \centerline{\includegraphics[width=.75\linewidth]{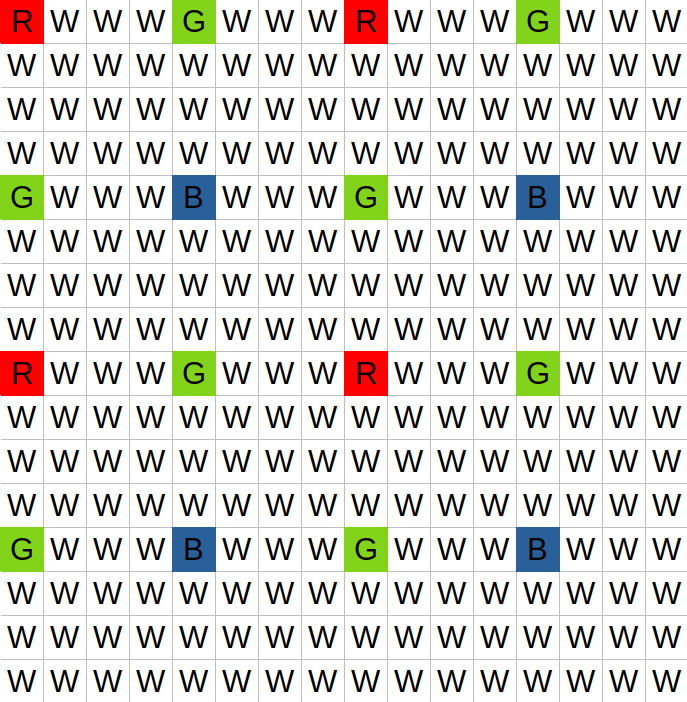}}
  \caption{Sparse 3 CFA}
\end{subfigure}

\medskip 
\begin{subfigure}{.475\linewidth}
  \centerline{\includegraphics[width=.75\linewidth]{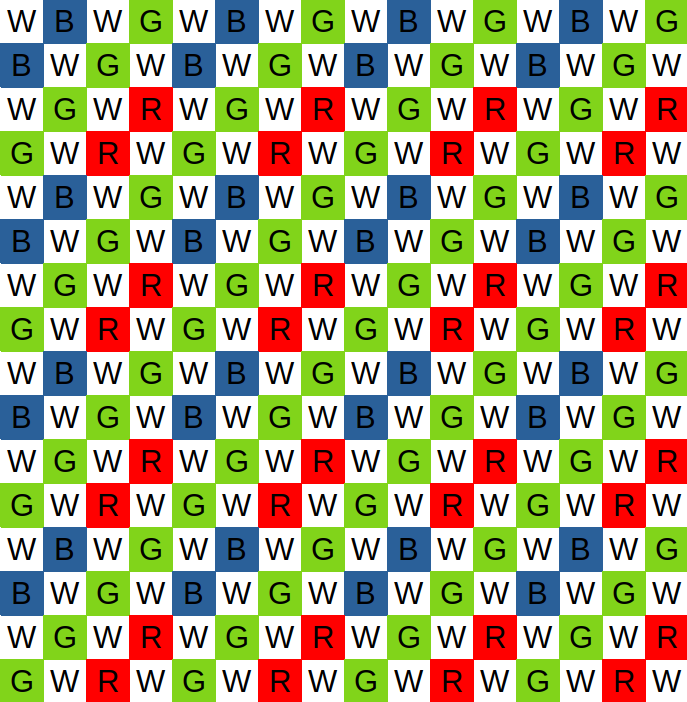}}
  \caption{Kodak CFA}
\end{subfigure}\hfill 
\begin{subfigure}{.475\linewidth}
  \centerline{\includegraphics[width=.75\linewidth]{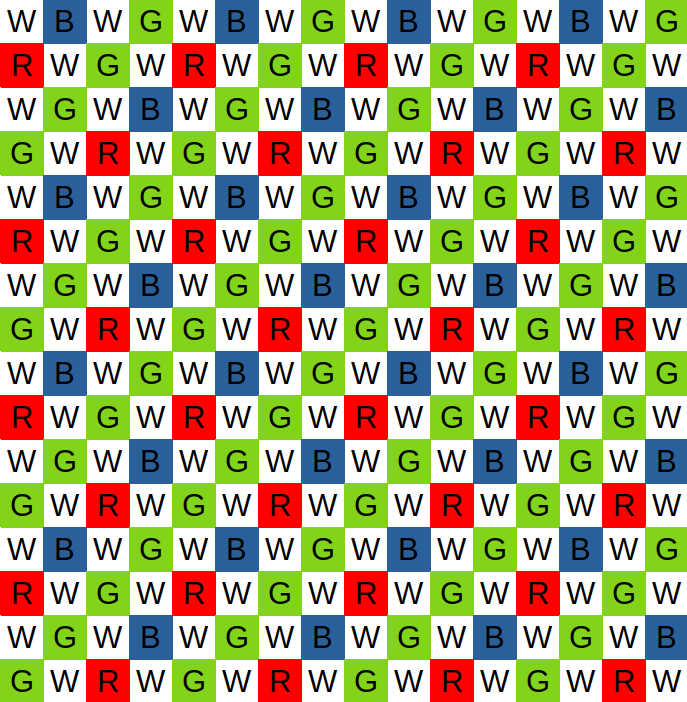}}
  \caption{Sony CFA}
\end{subfigure}

\caption{\small Bayer CFA and the three RGBW CFAs tested in the experimental section.}
\label{fig:patterns}
\end{figure}

The usual image acquisition process uses a predetermined CFA pattern to assign a color to each pixel of the sensor. While the most common pattern is the Bayer pattern Fig.~\ref{fig:bayer} \cite{cubic_spline, chrominance} other configurations exist. Some alternative designs exist, in which the filters' spectral response does not necessarily cover the RGB wavelengths. Such patterns are known with the more generic term spectral filter arrays (SFAs). For example, this category includes filters in the near infrared/ultra violet wavelength range, alternative color patterns such as cyan/magenta/yellow (CMY), and most relevantly to our work, wideband pixels. In the latter case, known as RGBW SFAs, the relevant sensor typically either does not include a filter, matching the sensor sensitivity response, or just cuts off the near infrared components. In layman terms, those pixels are often just referred as "white", and they allow for higher throughput of light energy which allows for increased brightness. An example of those patterns is the Sparse 3 CFA which is used in the AMICal Sat project \cite{amicalsat} to study northern lights from a microsatellite. Same goes for the Kodak and Sony patterns which are presented in Fig.~\ref{fig:patterns}.

The process of reconstructing the full color image from acquisitions taken with CFA-based cameras is known as demosaicking. As previously mentioned, this is typically done on the ground segment. It can be done in multiple ways: by classic interpolation between the pixels \cite{cubic_spline, linear}, by model-based methods which relies on a known mathematical model linking the input to the raw acquisition. More recently neural networks strategies appeared \cite{pipnet, random_network}, in general they are based on end-to-end training with large data-sets. However, all those methods have drawbacks. The interpolation methods often only work with specific patterns, posing restrictions for our scenario. Model-based approaches need a precise knowledge of the real acquisition conditions. We model the image formation with a mathematical formulation. This simulates the camera's acquisition process, inspired by \cite{daniele}. While allowing for remarkable flexibility, the model is very sensitive to inaccurate modeling of the real operations performed by the camera. Finally, data-driven based methods need a huge amount of data to be trained on and are often specialised on specific CFAs, leaving few possibilities for adaptation beside new training. For example, \cite{pipnet} proposes an efficient demosaicking neural network for Bayer CFA, but fails at reconstructing RGBW images. So is the case for the methods that focus only on Bayer CFA, like in \cite{linear, chrominance, cubic_spline}.

For those reasons this work proposes a general method able to solve the demosaicking problem for commercial designs of RGBW CFAs. This method relies on a model-based approach and minimizes a regularized convex cost function.

The novel contribution of this work are:

\begin{itemize}
    \item a forward model capable of simulating the camera's action, with support for the RGBW Sparse 3, Kodak and Sony patterns;
    \item a model-based method using Chambolle-Pock's algorithm to minimize a convex optimization problem solving the demosaicking challenge for RGBW patterns.
\end{itemize}

\section{Proposed framework}
\label{sec:proposed_method}

This section presents the acquisition process and the reconstruction. The first step is to model the image acquisition process usually done by the camera itself. This operation is performed by a forward model taking in input the reference image, in order to obtain the raw image. Then the reference image will be reconstructed into a RGB image from the raw image using a model-based method.

In the RGBW case the reference image, representing the incoming electromagnetic radiation from a scene under target is denoted by $\bm{X} \in \mathbb{R}^{M \times N \times (C + 1)}$ where $C$ is the number of channels ($C = 3$ for a RGB image), an additional channel is added containing the panchromatic, or "white", information. $M$ and $N$ being the spatial dimensions. The raw image representing the observation is encoded by $\bm{Y} \in \mathbb{R}^{M \times N}$, a scalar matrix coming from the forward model. Finally we denote by $\bm{\hat{X}} \in \mathbb{R}^{M \times N \times C}$ the image reconstructed from the raw image $\bm{Y}$.

\subsection{Forward model}
\label{ssec:forward_model}

We describe here our proposed framework for image formation, by reinterpreting the model proposed in \cite{daniele} for our target. In order to apply a CFA a mask $\bm{H} \in \mathbb{R}^{M \times N \times (C + 1)}$ is used, encoding the wanted CFA. The output of the CFA operation is obtained with:

\begin{equation}
    \bm{Y} = \sum_{k = 1}^{C + 1} \bm{X}_{k} \odot \bm{H}_{k} + \bm{N}
    \label{eq:forward_model}
\end{equation}

\noindent where $\odot$ is the Hadamard product between matrices, $\bm{X}_{k}$ and $\bm{H}_{k}$ are matrices which denote the $k$-th channel of the tensors $\bm{X}$ and $\bm{H}$ respectively. Moreover $\bm{N} \in \mathbb{R}^{M \times N}$ is a noise tensor representing the physical noise on the sensor. Additionally it models the mismatch between the forward model and the real camera. The tensor $\bm{H}$ encoding the mask contains a ones in the spectral channel associated to the corresponding spectral filter is kept. The rest is set at zeros, discarding the information at those positions. The image $\bm{Y}$ is then composed of scalar values, where for each pixel is kept only the information from the selected channels.

The deterministic part of eq.~\eqref{eq:forward_model} is a linear operation, which we denote as $A: \mathbb{R}^{M \times N \times (C + 1)} \rightarrow \mathbb{R}^{M \times N}$. We can therefore express $\bm{Y}$ as:

\begin{equation*}
    \bm{Y} = A(\bm{X}) + \bm{N}.
\end{equation*}

For convenience, we also derive here the adjoint $A^T(\bm{Y})$ of $A$ as the tensor $\bm{Z} \in \mathbb{R}^{M \times N \times (C + 1)}$ whose $k$-th channel is \cite{daniele}:

\begin{equation*}
    \bm{Z}_{k} = \bm{H}_{k} \odot \bm{Y} \;, \; \forall k \in \{ 1, ..., C + 1 \}.
\end{equation*}

\subsection{Model-based demosaicking}
\label{ssec:iterative_demo}

The demosaicking problem aims to reconstruct a RGB image from the raw acquisition. However this task is ill-posed as there are more unknowns than there is information as the goal is to recover data lost during the acquisition process. This problem can be reformulated as a minimization of a cost function with a regularization term:

\begin{equation}
    \bm{\hat{X}} = \underset{\bm{X}}{\arg\min} \left( \| A(\bm{X}) - \bm{Y} \|^2_F + \lambda \| L(\bm{X}) \|_{221} \right)
    \label{eq:problem}
\end{equation}

\noindent where $\| \cdot \|_F$ is the Frobenius norm, $\lambda \| L (\cdot) \|_{221}$ is the regularizer: $L$ is the isotropic total variation operator, $\lambda$ is a tuning parameter, and $||\cdot||_{221}$ denotes an $\ell_2$ norm in the channel and gradient domains followed by an $\ell_1$ norm in the spatial one. This term is derived from \cite{daniele} and \cite{totalvariation}, where experimentation yielded good results in a similar setup to ours. Note that in this setting $\bm{\hat{X}} \in \mathbb{R}^{M \times N \times (C + 1)}$, but as the method's goal is to reconstruct only the RGB image the panchromatic channel is simply discarded and $\bm{\hat{X}}$ is considered as a tensor of $\mathbb{R}^{M \times N \times C}$.

The Chambolle-Pock algorithm \cite{Esser_2010} is a well-known iterative algorithm used to solve optimization problems represented by the sum of two convex functions. This is the case of the demosaicking problem and it can be re-written in the Chambolle-Pock's framework:

\begin{equation*}
    \begin{array}{ll}
        \underset{\bm{X}}{\min} & f(\bm{X}) + g(L(\bm{X}))
    \end{array}
\end{equation*}

\noindent where $f$ and $g$ denote respectively the fidelity term $\| A (\cdot) - \bm{Y} \|^2_F$ and the regularization term $\lambda \| \cdot \|_{221}$ of the previous equation. The primal and dual variables are then updated from an arbitrary initialization with the following iterations \cite{Pock_2011, Esser_2010}:

\begin{equation}
    \left\{
    \begin{array}{ll}
        \bm{X}^{q + 1} & = \mathbf{prox}_{\tau, f}(\bm{X}^q - \tau L^T(\bm{Z}^q)) \\
        \bm{Z}^{q + 1} & = \mathbf{prox}_{\sigma, g^*}(\bm{Z}^q + \sigma L(2 \bm{X}^{q + 1} - \bm{X}^q))
    \end{array}
    \right.
    \label{eq:update_variables}
\end{equation}

\noindent with $\tau$ and $\sigma$ being the proximal steps for $f$ and $g^*$ (the Fenchel conjugate of $g$) respectively. $q$ is the iteration number which goes from $1$ to $Q$, the total number of iterations, $\bm{X}^q$ and $\bm{X}^q$ are the primal and dual variables at the $q$-th iteration.

\section{Experimental results}
\label{sec:results}

In this section we present the results on the RGBW patterns Sparse 3, Kodak and Sony. The associated code to reproduce the results of this section is publicly available on GitHub~\footnote{At https://github.com/mattmull42/RGBW\_demo}. We compare our method with the technique presented in \cite{e2v}, which is used on the microsatellite AMICal Sat and was quickly extended to work with Kodak and Sony patterns. The idea of the approach is to extract in one image the luminance (a scalar image with full spatial resolution) of the reference image by interpolating the pixels in the W channel associated with the RGB spectral filters. The colors are extracted in another image (of lower spatial resolution by full spectral resolution), by interpolating the R, G and B channels. Then those two images will be merged into a RGB image of full spectral and spatial resolution.

Concerning our method we need to tune the parameters introduced in eq.~\eqref{eq:problem} and eq.~\eqref{eq:update_variables} which are: $\lambda$, $\tau$, $\sigma$ and $Q$. The algorithm will run through $Q = 400$ iteration as empirical results showed no real improvement with higher number of iterations. The parameters $\lambda$, $\tau$ and $\sigma$ are optimized by the Python module Optuna \cite{Akiba_2019}.

\begin{figure}[!t]

\begin{minipage}[b]{1.0\linewidth}
  \centering
  \centerline{\includegraphics[width=6cm]{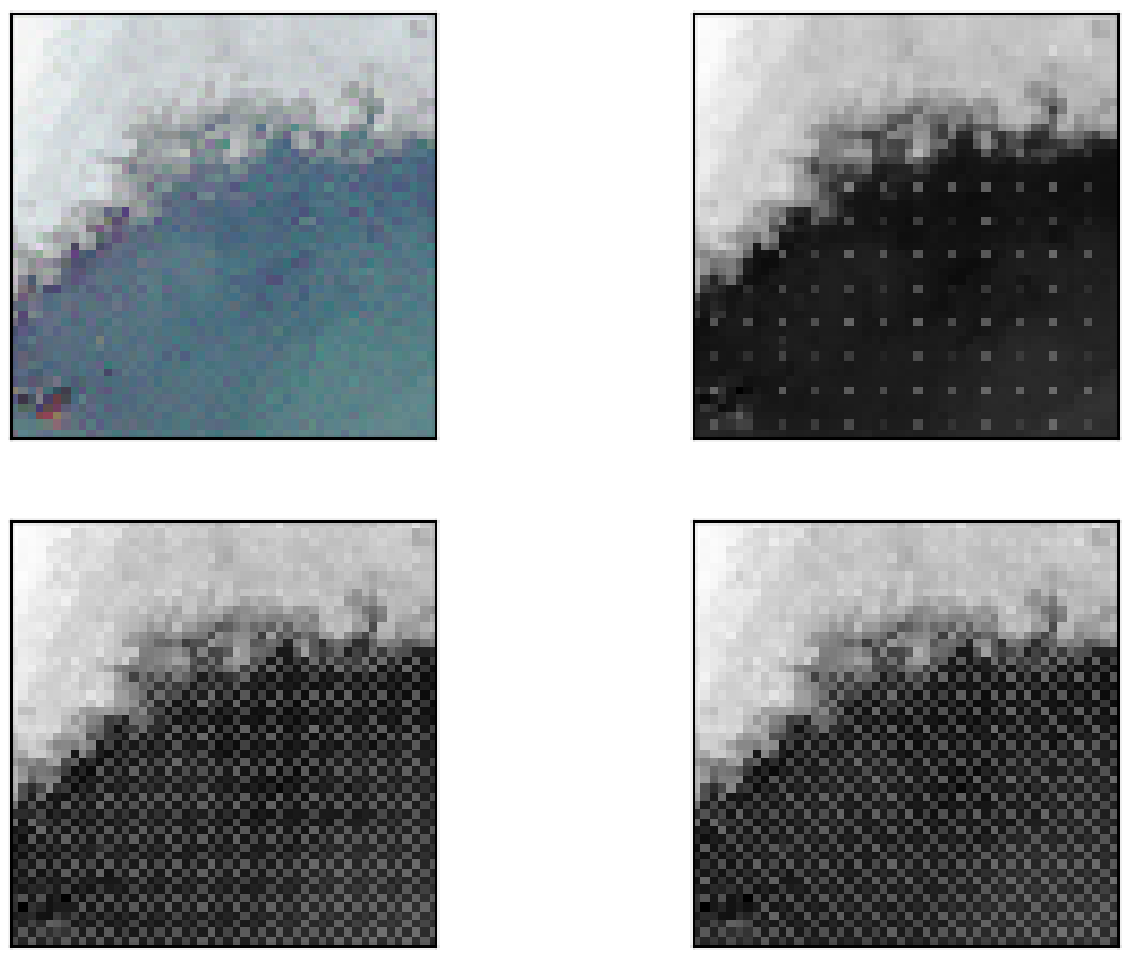}}
\end{minipage}
\caption{\small Example acquisitions with RGBW SFAs cameras. Reference scene (top left), outputs: Sparse 3 (top right), Kodak (bottom left) and Sony (bottom right).}
\label{fig:forward_results}
\end{figure}

The table of errors and the following figures are computed on the PAirMax remote sensing multispectral dataset \cite{9447896}. All the experiences are done with simulated acquisitions using our simulated forward model. The reference is not used in our reconstruction algorithm. We present in Fig.~\ref{fig:forward_results} a zoomed region of the outputs of the forward operator. The CFA patterns are noticeable, especially in the blue region.

\begin{figure}[!t]

\begin{minipage}[b]{1.0\linewidth}
  \centering
  \centerline{\includegraphics[width=5.6cm]{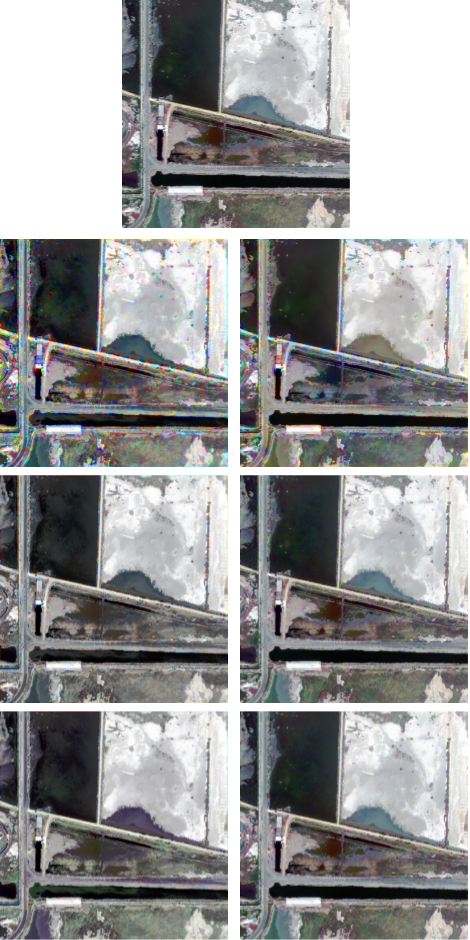}}
\end{minipage}
\caption{\small Reconstruction results. Reference is on top, baseline method on first column, proposed on second column. 2nd line is Sparse 3, 3rd line is Kodak and 4th line is Sony.}
\label{fig:results}
\end{figure}

Fig.~\ref{fig:results} shows the proposed reconstruction along with the pansharpening method on the three CFAs.

Table~\ref{tab:results} presents the mean and standard deviation of the mean squared error (MSE) of the reconstructions compared to the references, with and without an additive Gaussian noise of zero mean and standard deviation of 0.05.

\setlength{\tabcolsep}{0.05cm}
\begin{table}[htb]
\footnotesize
\centering
\begin{tabular}{|c|ll|ll|}
\hline
\multicolumn{1}{|l|}{\multirow{2}{*}{}} & \multicolumn{2}{c|}{Baseline \cite{e2v}}                                                                                                                                                                            & \multicolumn{2}{c|}{Proposed}                                                                                                                                                                                           \\ \cline{2-5} 
\multicolumn{1}{|l|}{}                  & \multicolumn{1}{c|}{Without noise}                                                                            & \multicolumn{1}{c|}{With noise}                                                          & \multicolumn{1}{c|}{Without noise}                                                                                 & \multicolumn{1}{c|}{With noise}                                                               \\ \hline
Sparse 3                                & \multicolumn{1}{l|}{\begin{tabular}[c]{@{}l@{}}$1.35 \times 10^{-2}$\\ $\pm 1.2 \times 10^{-2}$\end{tabular}} & \begin{tabular}[c]{@{}l@{}}$1.84 \times 10^{-2}$\\ $\pm 1.1 \times 10^{-2}$\end{tabular} & \multicolumn{1}{l|}{\begin{tabular}[c]{@{}l@{}}$\bm{1.95 \times 10^{-3}}$\\ $\pm 1.9 \times 10^{-3}$\end{tabular}} & \begin{tabular}[c]{@{}l@{}}$\bm{3.05 \times 10^{-3}}$\\ $\pm 2.5 \times 10^{-3}$\end{tabular} \\ \hline
Kodak                                   & \multicolumn{1}{l|}{\begin{tabular}[c]{@{}l@{}}$1.29 \times 10^{-2}$\\ $\pm 1.1 \times 10^{-2}$\end{tabular}} & \begin{tabular}[c]{@{}l@{}}$1.47 \times 10^{-2}$\\ $\pm 1.1 \times 10^{-2}$\end{tabular} & \multicolumn{1}{l|}{\begin{tabular}[c]{@{}l@{}}$\bm{2.63 \times 10^{-3}}$\\ $\pm 3.3 \times 10^{-3}$\end{tabular}} & \begin{tabular}[c]{@{}l@{}}$\bm{2.11 \times 10^{-3}}$\\ $\pm 2.8 \times 10^{-3}$\end{tabular} \\ \hline
Sony                                    & \multicolumn{1}{l|}{\begin{tabular}[c]{@{}l@{}}$1.24 \times 10^{-2}$\\ $\pm 1.1 \times 10^{-2}$\end{tabular}} & \begin{tabular}[c]{@{}l@{}}$1.40 \times 10^{-2}$\\ $\pm 1.1 \times 10^{-2}$\end{tabular} & \multicolumn{1}{l|}{\begin{tabular}[c]{@{}l@{}}$\bm{2.56 \times 10^{-3}}$\\ $\pm 3.2 \times 10^{-3}$\end{tabular}} & \begin{tabular}[c]{@{}l@{}}$\bm{2.82 \times 10^{-3}}$\\ $\pm 2.6 \times 10^{-3}$\end{tabular} \\ \hline
\end{tabular}
\caption{\small The mean and standard deviation of the MSE of the two methods, with and without noise (best results in bold).}
\label{tab:results}
\end{table}
\normalsize

The pansharpening method has the advantage to be quick to implement and to perform, as the results are obtained in less than a second. However it has two main drawbacks. Firstly it is dependent from the setup as it cannot be used in with a CFA where we do not have a majority of pixels without filter. Moreover this algorithm is dependent on the quality of the interpolations inside the pansharpening process.

On the contrary the proposed method is flexible and running it for $400$ iterations is only 20 seconds. It solves problems represented by eq.~\eqref{eq:problem}, and only the operator $A$ has to be adjusted to reflect a change in the acquisition setup. Our proposed algorithm can hence be applied almost effortlessly on CFA patterns such as the ones presented in this work.

\section{Conclusion}
\label{sec:conclusion}

In this work, we proposed a flexible image formation model for cameras that implement CFAs with RGBW patterns. This framework can be easily adjusted to adapt to different patterns. We proved that our proposed inversion method, which employs a Chambolle-Pock solver with a TV prior, improves the quality of the image by a factor of 10 in comparison to traditional interpolation techniques.


\bibliographystyle{IEEEbib}
\bibliography{refs}

\end{document}